\documentclass[onecolumn,showpacs,floatfix,aps,nofootinbib]{revtex4}
\usepackage[dvips]{graphics,color}
\usepackage{epsfig}\usepackage{float}
\RequirePackage{amssymb}
\usepackage{makeidx}
\usepackage[brazil]{babel}
\usepackage{graphicx}
\usepackage{indentfirst}
\usepackage{color}
\newcommand{\be}{\begin{equation}}
\newcommand{\ee}{\end{equation}}

\newcommand{\ba}{\begin{eqnarray}}
\newcommand{\ea}{\end{eqnarray}}

\begin{document}

\title{Torsion Effects in Braneworld Scenarios}
\author{J. M. Hoff da Silva}
\email{hoff@ift.unesp.br} \affiliation{}
\author{R. da Rocha}
\email{roldao.rocha@ufabc.edu.br} \affiliation{ Centro de
Matem\'atica, Computa\c c\~ao e Cogni\c c\~ao, Universidade
Federal do ABC, 09210-170, Santo Andr\'e, SP, Brazil}


\begin{abstract}
We present gravitational aspects of braneworld models endowed with
torsion terms both in the bulk and on the brane. In order to
investigate a conceivable and measurable gravitational effect, arising genuinely
from bulk torsion terms, we analyze the variation in the black hole
area by the presence of torsion. Furthermore, we extend the
well known results about consistency conditions in a framework that incorporates brane
torsion terms. It is shown, in a rough estimate, that the resulting effects
are generally suppressed by the internal space volume.
This formalism provides manageable models and their possible ramifications into
some aspects of gravity in this context, and cognizable corrections and physical effects as well.
\end{abstract}
\maketitle

\section{Introduction}

The presence of extra dimensions seems to be an almost inescapable
characteristic of high-energy physics based upon the auspices of
string theory. In this context, specific string theory inspired
scenarios, in which our universe is modeled by a brane --- the
braneworld scenario --- acquired special attention \cite{RSI} due
to the possibility of solving the hierarchy problem.
Concomitantly, the presence of torsion is also an output of string
theory \cite{cordas}. Indeed, when gravitation is recovered from
string theory, a plenty of physical fields abound, including the
torsion field. In this context, among other motivations, it seems
natural to explore some properties of braneworld models in the
presence of torsion.

The easiest method of introducing torsion terms in the theory is via
the addition of an antisymmetric part in the affine connection. The torsion tensor
components can be written in terms of the connection components
$\Gamma^{\rho}{}_{\beta\alpha}$ as $T^{\rho}{}_{\alpha\beta} =
\Gamma^{\rho}{}_{\beta\alpha} - \Gamma^{\rho}{}_{\alpha\beta}$.
The general connection components are related to the Levi-Civita
connection components $\mathring{\Gamma}{}^{\rho}{}_{\alpha\beta}$
--- associated with the spacetime metric $g_{\alpha\beta}$
components
--- through $\Gamma^{\rho}{}_{\alpha\beta} =
\mathring{\Gamma}{}^{\rho}{}_{\alpha\beta} +
K^{\rho}{}_{\alpha\beta}$, where $K^{\rho}{}_{\alpha \beta} =
\textstyle{\frac{1}{2}} \left( T_{\alpha}{}^{\rho}{}_{\beta} +
T_{\beta}{}^{\rho}{}_{\alpha} - T^{\rho}{}_{\alpha \beta} \right)$
denotes the contortion tensor components. Hereon the quantities
denoted by $\mathring{X}$ are constructed with the usual metric
compatible torsionless Levi-Civita connection components
$\mathring{\Gamma}{}^{\rho}{}_{\alpha\beta}$. We remark that the
source of contortion may be considered as the rank-2 antisymmetric
potential Kalb-Ramond (KR) field $B_{\alpha\beta}$, arising as a
massless mode in heterotic string theories \cite{cordas, SEN}.
Hereon we shall consider the formal geometric contortion, although
the contortion induced by the KR field can be considered in the
5-dimensional formalism when the prescription
$K^\rho_{\;\,\alpha\beta} = -
\frac{1}{M^{3/2}}H^\rho_{\;\,\alpha\beta}$ is taken into account,
where $H_{\rho\alpha\beta} = \partial_{[\rho}B_{\alpha\beta]}$ and
$M$ denotes the 5-dimensional Planck mass. The identification
between the KR field and the contortion can be always taken into
account when necessary, depending on the physical aspect of the
formalism that must be emphasized, although the formalism is not
precisely concerned with the fount of contortion, but with its
consequencese.

The combination of braneworld ideas encompassing torsion terms
seems to be an interesting approach in trying to ascertain new
signatures coming from high-energy physics. A complete scenario,
however, lacks, and we propose to delve into such a framework. In
this sense, some subterfuges are needed in order to extract
physical information from the model dealt with. The main aim of
this paper is to apply two well known strategies commonly used in
braneworld scenarios, in the context of models involving torsion:
the sum rules, bringing information about the brane torsion terms
behavior, and the Taylor expansion outside a black hole metric,
which gives information about the bulk torsion terms, where the
corrections in the area of the 5$D$ black string horizon are
evinced. In order to find some typical gravitational signatures of
braneworld scenarios with torsion we obtain all the formul\ae\,
for a Taylor expansion outside a black hole in Section II,
extending some results of Ref. \cite{Maartens} in order to
encompass accrued torsion corrections. It is shown how the
contortion and its derivatives determine the variation in the area
of the black hole horizon along the extra dimension, inducing
observable physical effects.

In  Section III we apply the braneworld consistency conditions in
the case when torsion is present in the brane manifold. We are
particularly concerned with the viability of such an extension,
analyzing the torsion effects in the brane scalar curvature. We
show that, for factorizable metrics, the torsion contribution to
the brane curvature is damped by the distance between the branes.
In warped braneworld models, however, this damping is --- at least
partially --- compensated by terms of the warp factor. It is also
shown that if the brane manifold is endowed with a connection
presenting torsion then a Randall-Sundrum like scenario with equal
sign brane tension becomes possible, in acute contrast to the
standard Randall-Sundrum model. Roughly speaking, this last
possibility comes from the following reasoning. The presence of
the torsion terms generally relax the consistence conditions,
specially in what concerns the sum over the brane tensions.
Therefore, the brane tensions are not restricted to the same sign,
although constrained by a specific contraction of contortion
terms.

Our program throughout this paper explicitly consists of the
following: the next Section deals with the question of finding a
possible departure from the usual (torsionless) braneworld
scenario, induced exclusively from bulk torsion terms within
braneworld scenarios endowed with a general affine connection
presenting torsion. Particularly, it is accomplished by looking
the variation in the black hole horizon due to an specific
arrangement of contortion terms. In Section III, we analyze some
aspects related to the feasibility of braneworld manageable models
with brane torsion terms. By studying the general consistency
conditions applied to this case, we arrive at some roughly
estimates concerning brane torsion effects. Although the presence
of torsion is not prohibited at all, its effects are generally
suppressed.

Our generalization concerns to construct a formal framework that
contains terms of contortion, capable --- for instance --- to
correct  black holes horizons and consequent measurement of
variation in quasars luminosities, as in \cite{merc} and
references therein. Also, the braneworld consistency conditions
are investigated in terms of contortion terms corrections
demonstrating that for factorizable metrics, the contortion
contribution to the brane curvature is damped by the distance
between the branes.

\section{Measurable Torsion Effects}

In a previous paper \cite{NOIS} we proved that although the
presence of torsion terms in the connection does not modify the
Israel-Darmois matching conditions, despite of the modification in
the extrinsic curvature and in the connection, the Einstein
equation, obtained using the Gauss-Codazzi formalism, is extended.
The factors involving contortion alter drastically the effective
Einstein equation on the brane, and also the function involving contortion
terms that is analogous to the effective
cosmological constant as well.

We shall use such results to extend the bulk metric Taylor expansion in terms of the brane metric, in a direction
orthogonal to the brane, encompassing torsion terms. As an
immediate application, the corrections in a black hole horizon
area due to contortion terms are achieved.

We have investigated the matching conditions in the presence of
torsion terms, and under the assumptions of discontinuity across
the brane, both of the junction conditions are shown to be the
same as the usual case. As the covariant derivative is modified by
the contortion, the extrinsic curvature is also modified, and the
conventional arguments point in the direction of some modification
in the matching conditions. However, it seems that the role of
torsion terms in the braneworld picture is restricted to the
geometric part of effective Einstein equation on the brane. More
explicitly, looking at the equation that relates the Einstein
equation in four dimensions with bulk quantities (see reference
\cite{JP}, for instance) it follows that
\begin{eqnarray}
\hspace{-0.5cm}^{(4)}\!G_{\rho\sigma}&=&\left.\frac{2k_{5}^{2}}{3}\Bigg(T_{\alpha\beta}q_{\rho}^{\;\alpha}q_{\sigma}^{\;\beta}+(T_{\alpha\beta}
n^{\alpha}n^{\beta}-\frac{1}{4}T)q_{\rho\sigma}\Bigg)+
\Xi\Xi_{\rho\sigma}-\Xi_{\rho}^{\;\alpha}\Xi_{\alpha\sigma}-
\frac{1}{2}q_{\rho\sigma}(\Xi^{2}-\Xi^{\alpha\beta}\Xi_{\alpha\beta})-\;^{(5)}\!C^{\alpha}_{\;\beta\gamma\epsilon}n_{\alpha}n^{\gamma}q_{\rho}^{\;\beta}q_{\sigma}^{\;\epsilon},\right.\label{gde}
\end{eqnarray}
where $T_{\rho\sigma}$ denotes the energy-momentum tensor,
$\Xi_{\rho\sigma}=q_{\rho}^{\;\alpha}q_{\sigma}^{\;\beta}\nabla_{\alpha}n_{\beta}$
is the extrinsic curvature, $k_5$ denotes the 5-dimensional
gravitational constant, and
$^{(5)}\!C^{\alpha}_{\;\beta\rho\sigma}$ denotes the Weyl tensor.
By restricting to quantities evaluated on the brane, or tending to
the brane, we see that the only way to get some contribution from
torsion terms is via the term $^{(4)}\!G_{\rho\sigma}$, and also
via the Weyl tensor. Supposing $\mathbb{Z}_{2}$-symmetry, the
extrinsic curvature reads
\ba\Xi^{+}_{\alpha\beta}=-\Xi^{-}_{\alpha\beta}=-2G_{N}\Big(\pi_{\alpha\beta}-\frac{q_{\alpha\beta}\pi^{\gamma}_{\gamma}}{4}\Big).\label{c1}\ea

Decomposing the stress-tensor associated with the bulk in
$T_{\alpha\beta}=-\Lambda g_{\alpha\beta}+\delta S_{\alpha\beta}$
and $S_{\alpha\beta}=-\lambda q_{\alpha\beta}+\pi_{\alpha\beta}$,
where $\Lambda$ is the bulk cosmological constant and $\lambda$
the brane tension, and substituting into Eq.(\ref{gde}) it follows
that \ba ^{(4)}\!G_{\mu\nu}=-\Lambda_{4}q_{\mu\nu}+8\pi
G_{N}\pi_{\mu\nu}+k_{5}^{4}Y_{\mu\nu}-E_{\mu\nu} ,\label{c3}\ea
where
$E_{\mu\nu}=^{(5)}\!C^{\alpha}_{\beta\gamma\sigma}n_{\alpha}n^{\gamma}q_{\mu}^{\beta}q_{\nu}^{\sigma}$
encodes the Weyl tensor contribution, $G_{N}=\frac{\lambda
k_{5}^{4}}{48\pi}$ is the analogue of the Newton gravitational
constant, the tensor $Y_{\mu\nu}$ is quadratic in the brane
stress-tensor and given by
$Y_{\mu\nu}=-\frac{1}{4}\pi_{\mu\alpha}\pi_{\nu}^{\alpha}+\frac{1}{12}\pi^{\gamma}_{\gamma}\pi_{\mu\nu}+\frac{1}{8}
q_{\mu\nu}\pi_{\alpha\beta}\pi^{\alpha\beta}-\frac{1}{2}q_{\mu\nu}(\pi^{\gamma}_{\gamma})^{2}$,
and
$\Lambda_{4}=\frac{k_{5}^{2}}{2}\Big(\Lambda+\frac{1}{6}k_{5}^{2}\lambda^{2}\Big)$
denotes the effective brane cosmological constant.

Using the Einstein tensor on the brane encoding torsion terms, the
$E_{\mu\nu}$ tensor can be expressed in terms of the bulk
contortion terms by
\begin{eqnarray}
\hspace{-.5cm}E_{\kappa\delta}&=&\left.\mathring{E}_{\kappa\delta}+\Big(\nabla_{[\nu}K^{\mu}_{\;\;\alpha\beta]}+
K^{\mu}_{\;\;\gamma[\nu}K^{\gamma}_{\;\;\alpha\beta]}\Big)n_{\mu}n^{\nu}
q_{\kappa}^{\alpha}q_{\delta}^{\beta}\right.-\frac{2}{3}(q_{\kappa}^{\alpha}q_{\delta}^{\beta}+n^{\alpha}n^{\beta}q_{\kappa\delta})
\Big(\nabla_{[\lambda}K^{\lambda}_{\;\;\beta\alpha]}+K^{\lambda}_{\;\;\gamma\lambda}K^{\gamma}_{\;\;\beta\alpha}
-K^{\sigma}_{\;\;\beta\gamma}K^{\gamma}_{\;\;\sigma\alpha}\Big)\nonumber\\&&+\frac{1}{6}q_{\kappa\delta}\left.
\Big(2\nabla^{\lambda}K^{\tau}_{\;\;\lambda\tau}-K_{\tau\lambda}^{\;\;\;\lambda}K^{\tau\gamma}_{\;\;\;\gamma}+K_{\tau\gamma\lambda}
K^{\tau\lambda\gamma} \Big), \right. \label{c8}
\end{eqnarray} where $\nabla_{\mu}$ is the bulk covariant
derivative. Now, the explicit influence of the contortion terms in
the Einstein brane equation can be visualized. From
Eqs.(\ref{c3}), (\ref{c8}) and expressing the torsion terms of the
Einstein brane tensor (see Eq. (20) of reference \cite{NOIS}), it
follows that
\begin{widetext}
\begin{eqnarray}&&
\left.
\hspace{-1cm}^{(4)}\!\mathring{G}_{\mu\nu}+D_{[\lambda}\;^{(4)}\!K^{\lambda}_{\;\;\mu\nu]}+
^{(4)}\!K^{\delta}_{\;\;\gamma\delta}\;^{(4)}\!K^{\lambda}_{\;\;\mu\nu}
-
\;^{(4)}\!K^{\sigma}_{\;\;\nu\gamma}\;^{(4)}\!K^{\gamma}_{\;\;\sigma\mu}
=-\tilde{\Lambda}_{4}q_{\mu\nu}+8\pi
G_{N}\pi_{\mu\nu}+k_{5}^{4}Y_{\mu\nu}-\mathring{E}_{\mu\nu}\right.\nonumber
\\&&\hspace{-1cm}+q_{\mu}^{\alpha}q_{\nu}^{\beta} \left.\Bigg[\frac{2}{3}\Big(\nabla_{[\lambda}
K^{\lambda}_{\;\;\beta\alpha]}+K^{\sigma}_{\;\;\gamma\sigma}K^{\gamma}_{\;\;\beta\alpha}
-K^{\lambda}_{\;\;\beta\gamma}K^{\gamma}_{\;\;\lambda\alpha}\Big)-n_{\rho}n^{\sigma}\Big(\nabla_{[\sigma}K^{\rho}_{\;\;\alpha\beta]}+K^{\rho}_{\;\;\gamma[\sigma}K^{\gamma}_{\;\;\alpha\beta]}
\Big)\Bigg], \right. \label{c9}
\end{eqnarray}
where
\begin{eqnarray}
\tilde{\Lambda}_{4}&\equiv&\left.
\Lambda_{4}-D^{\lambda}\;^{(4)}\!K^{\tau}_{\;\;\lambda\tau}+\frac{1}{2}\;^{(4)}K_{\tau\alpha}^{\;\;\;\alpha}\;^{(4)}K^{\tau\lambda}_{\;\;\;\lambda}
-\frac{1}{2}\;\;^{(4)}K_{\tau\gamma\lambda}\;^{(4)}K^{\tau\lambda\gamma}-\frac{2}{3}n^{\alpha}n^{\beta}\Big(\nabla_{\lambda}K^{\lambda}_{\;\;\beta\alpha}-
\nabla_{\alpha}K^{\lambda}_{\;\;\beta\lambda}\right.\nonumber\\&&+\left.K^{\lambda}_{\;\;\gamma\lambda}K^{\gamma}_{\;\;\beta\alpha}
-K^{\sigma}_{\;\;\beta\gamma}K^{\gamma}_{\;\;\sigma\alpha}\Big)+\frac{1}{6}\Big(2\nabla^{\lambda}K^{\tau}_{\;\;\lambda\tau}-
K_{\tau\alpha}^{\;\;\;\alpha}K^{\tau\lambda}_{\;\;\;\lambda}+K_{\tau\gamma\lambda}K^{\tau\lambda\gamma}\Big).\label{c10}
\right.
\end{eqnarray}
\end{widetext} The function $\tilde{\Lambda}_4$ above is usually called
 effective cosmological ``constant'' in the literature, in the sense that all its terms are multiplied by the brane metric
 in the Einstein effective equation (\ref{c9}). Eqs. (\ref{c9}) and (\ref{c10}) show that the factors involving
contortion, both in four and in five dimensions, modify
drastically the effective Einstein equation on the brane and the
effective cosmological constant as well.

Now, let us look at some deviations of the black hole horizon
coming from the bulk torsion terms. Hereon in this Section we
assume vacuum on the brane ($\pi_{\mu\nu}=0=Y_{\mu\nu}$) and
neglect the contribution of the effective cosmological constant
term, which is expected to be smaller, by some orders of
magnitude, than the contribution of the term Weyl \cite{Maartens}.
Using a Taylor expansion in the extra dimension in order to probe
properties of a static black hole on the brane \cite{Dad}, the
bulk metric can be written as
\begin{eqnarray}\label{metrica}
g_{\mu\nu}(x,y) &=& q_{\mu\nu} - (\mathring{E}_{\mu\nu} +
A_{\mu\nu} )y^2 - \frac{2}{l}(\mathring{E}_{\mu\nu} + A_{\mu\nu})
y^3 +\frac{1}{12}\left.\Bigg(\left({\Box}\mathring{E}_{\mu\nu} -
\frac{32}{l^2}\mathring{E}_{\mu\nu} +
2\mathring{R}_{\mu\alpha\nu\beta}\mathring{E}^{\alpha\beta} +
6\mathring{E}_{\mu}^{\;
\alpha}\mathring{E}_{\alpha\nu}\right)\right.\nonumber\\&&+
\left.\left({\Box}{A}_{\mu\nu} - \frac{32}{l^2}{A}_{\mu\nu} + 2(
\nabla_{[\nu}K_{\mu\alpha\beta]}){A}^{\alpha\beta} +
2K_{\mu\gamma[\nu}K^{\gamma}_{\;\;\alpha\beta]}{A}^{\alpha\beta} +
6\mathring{A}_{\mu}^{\; \alpha}\mathring{A}_{\alpha\nu}\right)
 \Bigg)\right. y^4 + \cdots
\nonumber\end{eqnarray} \noindent where \begin{eqnarray}A_{\mu\nu}
&=&  \left.\Big(\nabla_{[\delta}K^{\kappa}_{\;\;\alpha\beta]}+
K^{\kappa}_{\;\;\gamma[\beta}K^{\gamma}_{\;\;\alpha\delta]}\Big)n_{\kappa}n^{\delta}
q_{\mu}^{\alpha}q_{\nu}^{\beta}\right. +\frac{1}{6}q_{\mu\nu}
\Big(2\nabla^{\lambda}K^{\tau}_{\;\;\lambda\tau}-K_{\tau\lambda}^{\;\;\;\lambda}K^{\tau\gamma}_{\;\;\;\gamma}+K_{\tau\gamma\lambda}
K^{\tau\lambda\gamma}
\Big)\nonumber\\&&\left.-\frac{2}{3}(q_{\mu}^{\alpha}q_{\nu}^{\beta}+n^{\alpha}n^{\beta}q_{\mu\nu})
\Big(\nabla_{[\lambda}K^{\lambda}_{\;\;\beta\alpha]}+K^{\lambda}_{\;\;\gamma\lambda}K^{\gamma}_{\;\;\beta\alpha}
-K^{\sigma}_{\;\;\beta\gamma}K^{\gamma}_{\;\;\sigma\alpha}\Big)\right.\nonumber\label{amunu}\end{eqnarray}
and $\Box$ denotes the usual d'Alembertian. As in \cite{Maartens},
it shows in particular that the propagating effect of $5D$ gravity
arises only at the fourth order of the expansion. For a static
spherical metric on the brane given by \begin{equation}\label{124}
g_{\mu\nu}dx^{\mu}dx^{\nu} = - F(r)dt^2 + \frac{dr^2}{H(r)} +
r^2d\Omega^2,
\end{equation}
\noindent
 the projected Weyl term on the brane is given by the expressions\footnote{In the three expressions below, the indices $r$ and $\theta$ strictly denote
 the coordinates, and can not be confounded with summation indices.}
 \begin{eqnarray}
E_{00} &=& \frac{F}{r}\left(H' - \frac{1 - H}{r}\right) +
\Big(\nabla_{\nu}K^{\mu 00}-\nabla_{0}K^{\mu}_{\;\;0\nu}+
K^{\mu}_{\;\;\gamma\nu}K^{\gamma}_{\;\;00}-K^{\mu}_{\;\;\gamma
0}K^{\gamma}_{\;\;0\nu}\Big)n_{\mu}n^{\nu}
 F^2\nonumber\\&&-\frac{2}{3} F(F-1)
\Big(\nabla_{\lambda}K^{\lambda}_{\;\;00}-\nabla_{0}K^{\lambda}_{\;\;0\lambda}+K^{\lambda}_{\;\;\gamma\lambda}K^{\gamma}_{\;\;00}
-K^{\sigma}_{\;\;0\gamma}K^{\gamma}_{\;\;\sigma 0}\Big)
+\frac{1}{6} F
\Big(2\nabla^{\lambda}K^{\tau}_{\;\;\lambda\tau}-K_{\tau\lambda}^{\;\;\;\lambda}K^{\tau\gamma}_{\;\;\;\gamma}+K_{\tau\gamma\lambda}
K^{\tau\lambda\gamma} \Big),\nonumber\\
E_{rr} &=& -\frac{1}{rH}\left(\frac{F'}{F} - \frac{1 -
H}{r}\right)  + \Big(\nabla_{\nu}K^{\mu
rr}-\nabla_{r}K^{\mu}_{\;\;r\nu}+
K^{\mu}_{\;\;\gamma\nu}K^{\gamma}_{\;\;rr}-K^{\mu}_{\;\;\gamma
r}K^{\gamma}_{\;\;r\nu}\Big)n^{\mu}n^{\nu}
 H^{-2}\nonumber\\&&-\frac{2}{3} H^{-1}(H^{-1}-(n^r)^2)
\Big(\nabla_{\lambda}K^{\lambda}_{\;\;rr}-\nabla_{r}K^{\lambda}_{\;\;r\lambda}+K^{\lambda}_{\;\;\gamma\lambda}K^{\gamma}_{\;\;rr}
-K^{\sigma}_{\;\;r\gamma}K^{\gamma}_{\;\;\sigma
r}\Big)\nonumber\\&& +\frac{1}{6} H^{-1}\left.
\Big(2\nabla^{\lambda}K^{\tau}_{\;\;\lambda\tau}-K_{\tau\lambda}^{\;\;\;\lambda}K^{\tau\gamma}_{\;\;\;\gamma}+K_{\tau\gamma\lambda}
K^{\tau\lambda\gamma} \Big) \right. ,\nonumber\\
E_{\theta\theta} &=& -1 + H +\frac{r}{2}H\left(\frac{F'}{F} +
\frac{H'}{H}\right) + \Big(\nabla_{\nu}K^{\mu
\theta\theta}-\nabla_{\theta}K^{\mu}_{\;\;\theta\nu}+
K^{\mu}_{\;\;\gamma\nu}K^{\gamma}_{\;\;\theta\theta}-K^{\mu}_{\;\;\gamma
\theta}K^{\gamma}_{\;\;\theta\nu}\Big)n_{\mu}n^{\nu}
 r^4\nonumber\\&&-\frac{2}{3} r^2(r^2+1)
\Big(\nabla_{\lambda}K^{\lambda}_{\;\;\theta\theta}-\nabla_{\theta}K^{\lambda}_{\;\;\theta\lambda}+K^{\lambda}_{\;\;\gamma\lambda}K^{\gamma}_{\;\;\theta\theta}
-K^{\sigma}_{\;\;\theta\gamma}K^{\gamma}_{\;\;\sigma \theta} -
\frac{1}{2} \nabla^{\lambda}K^{\tau}_{\;\;\lambda\tau}+\frac{1}{4}
K_{\tau\lambda}^{\;\;\;\lambda}K^{\tau\gamma}_{\;\;\;\gamma}-\frac{1}{4}K_{\tau\gamma\lambda}
K^{\tau\lambda\gamma} \Big).\label{1333}
\end{eqnarray}

\noindent Note that in Eq.(\ref{124}) the metric reduces to the
Schwarzschild one, if $F(r)$ equals $H(r)$. The exact
determination of these radial functions remains an open problem in
black hole theory on the brane \cite{Maartens}.

These components allow one to evaluate the metric coefficients in
Eq.(\ref{metrica}). The area of the $5D$ horizon is determined by
$g_{\theta\theta}$. Defining $\psi(r)$ as the deviation from a
Schwarzschild form $H$, i.e.,
\begin{equation}\label{h}
H(r) = 1 - \frac{2M}{r} + \psi(r),
\end{equation}
\noindent where $M$ is constant, yields
\begin{eqnarray}\label{gtheta}
g_{\theta\theta}(r,y) &=& r^2  + \psi'\left(1 +
\frac{2}{l}y\right) + \Big(\nabla_{\nu}K^{\mu
\theta\theta}-\nabla_{\theta}K^{\mu}_{\;\;\theta\nu}+
K^{\mu}_{\;\;\gamma\nu}K^{\gamma}_{\;\;\theta\theta}-K^{\mu}_{\;\;\gamma
\theta}K^{\gamma}_{\;\;\theta\nu}\Big)n_{\mu}n^{\nu}
 r^4\nonumber\\&-&\frac{2}{3} r^2(r^2+1)
\Big(\nabla_{[\lambda}K^{\lambda}_{\;\;\theta\theta]}+K^{\lambda}_{\;\;\gamma\lambda}K^{\gamma}_{\;\;\theta\theta}
-K^{\sigma}_{\;\;\theta\gamma}K^{\gamma}_{\;\;\sigma \theta}
-\frac{1}{2}
\nabla^{\lambda}K^{\tau}_{\;\;\lambda\tau}+\frac{1}{4}K_{\tau\lambda}^{\;\;\;\lambda}K^{\tau\gamma}_{\;\;\;\gamma}-\frac{1}{4}K_{\tau\gamma\lambda}
K^{\tau\lambda\gamma} \Big)y^2\nonumber\\ &+&\cdots \label{modif}
\end{eqnarray}
\noindent It shows how $\psi$ \emph{and} the contortion and its
derivatives determine the variation in the area of the horizon along
the extra dimension. Also, the variation in the black string
properties can be extracted. Obviously, when the torsion goes to
zero, all the results above are led to the ones obtained in
\cite{Maartens}, \cite{Dad}, and references therein.
 In particular, Eq.(\ref{metrica}) --- when the torsion, and consequently $A_{\mu\nu}$ defined in Eq.(\ref{amunu}),  goes to zero ---
 is led to the results previously obtained in \cite{Maartens}.

As the area of the a black hole $5D$ horizon is determined by
$g_{\theta\theta}$, in particular it may indicate observable
signatures of corrections induced by contortion terms, since for a
given fixed effective extra dimension size, supermassive black
holes give the upper limit of variation in luminosity of quasars.
Also, it is possible to re-analyze how the quasar luminosity
variation behaves as a function of the AdS$_5$ bulk radius ---
corrected by contortion terms --- in some solar mass range, as in
\cite{merc} and references therein.

Furthermore,  braneworld measurable corrections induced by
contortion terms for  quasars, associated with Schwarzschild and
Kerr black holes, by their luminosity observation are important.
These corrections in a torsionless context were shown to be more
notorious for mini-black holes, where the Reissner-Nordstrom
radius in a braneworld scenario is shown to be around a hundred
times bigger than the standard ReissnerNordstrom radius
associated with mini-black holes, besides mini-black holes being much more
sensitive to braneworld effects. It is possible to repeat all the
comprehensive and computational procedure in \cite{merc} in order
to verify how the contortion effects in Eq.(\ref{modif}) can
modify even more the above-mentioned results. In addition, such
corrections involving torsion can also affect the mini-black holes radii
(horizons), considering braneworld effects in ADD and
Randall-Sundrum models, as presented in \cite{merc}, where the
radius of a black holes on the brane is much smaller than the size scale of
the extra dimensions, and the black hole can be well described by
the classical solutions of higher-dimensional Einstein equations.
The Schwarzschild radius, in the context of Myers-Perry
extra-dimensional formalism, is shown to be significantly
increased by tidal charge and spinning effects and can be also
corrected by contortion terms, as presented in Eq.(\ref{modif}).
Such corrections may give more precise calculations concerning
cross sections, Planck and mini-black holes masses, and Hawking
temperature, contributing in this way to a more complete, precise,
and realistic analysis of mini-black holes production in the next
generation of particle colliders, such as LHC. For details in an
torsionless scenario see \cite{merc} and references therein.

Finally, the applications of Eq.(\ref{modif}) are not the main aim
here, and shall be addressed in a forthcoming publication,
although such applications evince the importance of
Eq.(\ref{modif}) in the context where any physical quantity
involving black string horizon can be deeper investigated in the
light of contortion corrections.

The modification in the area of the black hole horizon due to
torsion terms, whose functional form is depicted in Eq.
(\ref{modif}), can be better appreciated in a specific basis, i.
e., an explicit {\it ansatz} for the spacetime metric. This is,
however, out of the scope of the present work. The important point
here is that torsion terms do affect the black hole horizon and
the departure from the usual (torsionless) case is precisely given
by Eq. (\ref{modif}). In the next Section we extend and apply the
braneworld sum rules to the case with torsion, considering some
estimates of the torsion effects.

\section{Sum rules with torsion}

In this Section we shall derive the consistency conditions for
braneworld scenarios embedded in a Riemann-Cartan manifold. The general
procedure is quite similar to the one found in \cite{RK,LE} and we
shall comprise some of the general formulation here, for the sake of
completeness.

We start in a very general setup, analyzing a $D$-dimensional bulk
spacetime geometry, endowed with a non-factorizable metric \ba
ds^{2}&=&\left.G_{AB}dX^{A}dX^{B}\right.\nonumber\\&=&
\left.W^{2}(r)g_{\alpha\beta}dx^{\alpha}dx^{\beta}+g_{ab}(r)dr^{a}dr^{b}\label{2},\right.
\ea where $W^{2}(r)$ is the warp factor, $X^{A}$ denotes the
coordinates of the full $D$-dimensional bulk, $x^{\alpha}$ stands
for the $(p+1)$ non-compact spacetime coordinates, and $r^{a}$
labels the $(D-p-1)$ directions in the internal compact space. The
$D$-dimensional Ricci tensor can be related to its lower
dimensional partners by \cite{RK} \begin{eqnarray}
R_{\mu\nu}&=&\bar{R}_{\mu\nu}-\frac{g_{\mu\nu}}{(p+1)W^{p-1}}\nabla^{2}W^{p+1},\label{21}
\\
R_{ab}&=&\tilde{R}_{ab}-\frac{p+1}{W}\nabla_{a}\nabla_{b}W,\label{3}\end{eqnarray}
where $\tilde{R}_{ab}$, $\nabla_{a}$ and $\nabla^{2}$ are
respectively the Ricci tensor, the covariant derivative, and the
Laplacian operator constructed by means of  the internal space
metric $g_{ab}$. $\bar{R}_{\mu\nu}$ is the Ricci tensor derived
from $g_{\mu\nu}$. Denoting the three curvature scalars by
$R=G^{AB}R_{AB}$, $\bar{R}=g^{\mu\nu}\bar{R}_{\mu\nu}$, and
$\tilde{R}=g^{ab}\tilde{R}_{ab}$ we have, from Eqs.(\ref{2}) and
(\ref{3}), \ba
\frac{1}{p+1}\Big(W^{-2}\bar{R}-R^{\mu}_{\;\,\mu}\Big)=pW^{-2}\nabla
W\cdot\nabla W+W^{-1}\nabla^{2}W \label{4}\ea and \be
\frac{1}{p+1}\Big(\tilde{R}-R_{\;\,a}^{a}\Big)=W^{-1}\nabla^{2}W,\label{5}
\ee where $R^{\mu}_{\;\,\mu}\equiv W^{-2}g^{\mu\nu}R_{\mu\nu}$ and
$R^{a}_{\;\,a}\equiv g^{ab}R_{ab}$
($R=R^{\mu}_{\;\,\mu}+R^{a}_{\;\,a}$). It can be easily verified
that for an arbitrary constant $\xi$ the following identity holds
\be \frac{\nabla \cdot (W^{\xi}\nabla W)}{W^{\xi+1}}=\xi
W^{-2}\nabla W\cdot \nabla W+W^{-1}\nabla^{2}W \label{6}.\ee
Combining the above equation with Eqs.(\ref{2}) and (\ref{3}) we
have \ba \nabla \cdot (W^{\xi}\nabla
W)=\frac{W^{\xi+1}}{p(p+1)}[\xi\big(W^{-2}\bar{R}-R^{\mu}_{\;\,\mu}\big)
+(p-\xi)\big(\tilde{R}-R^{a}_{\;\,a}\big)].\label{7} \ea

The $D$-dimensional Einstein equation is given by \be R_{AB}=8\pi
G_{D}\Big(T_{AB}-\frac{1}{D-2}G_{AB}T\Big),\label{8} \ee where
$G_{D}$ is the gravitational constant in $D$ dimensions. It is
easy to write down the following equations: \ba
R^{\mu}_{\;\,\mu}=\frac{8\pi
G_{D}}{D-2}(T^{\mu}_{\;\,\mu}(D-p-3)-T^{m}_{\;\,m}(p+1)),\quad\qquad
R^{m}_{\;\,m}=\frac{8\pi
G_{D}}{D-2}(T^{m}_{\;\,m}(p-1)-T^{\mu}_{\;\,\mu}(D-p-1)).\label{9}\ea
In the above equations we set
$T^{\mu}_{\;\,\mu}=W^{-2}g_{\mu\nu}T^{\mu\nu}$
($T^{M}_{\;\,M}=T^{\mu}_{\;\,\mu}+T^{m}_{\;\,m}$). Now, it is
possible to relate $R^{\mu}_{\;\,\mu}$ and $R^{m}_{\;\,m}$ in
Eq.(\ref{7}) in terms of the stress-tensor. Note that the left
hand side of  Eq.(\ref{7}) vanishes upon integration
along a compact internal space. Hence, taking all that into
account we have \ba  \oint
W^{\xi+1}\Bigg(T^{\mu}_{\;\,\mu}[(p-2\xi)(D-p-1)+2\xi]+T^{m}_{\;\,m}p\,(2\xi-p+1)+\frac{D-2}{8\pi
G_{D}}[(p-\xi)\tilde{R}+\xi \bar{R}W^{-2}]\Bigg)=0.\label{11} \ea

Heretofore in this Section we have just reproduced the well
established results on consistency conditions applied to
braneworld scenarios. Going further, the last equation --- which provides a
one parameter family of consistency conditions for warped
braneworld scenarios in arbitrary dimensions --- is applied to the case
when torsion terms are present on the brane. First, however, we
want to particularize the analysis for a 5-dimensional bulk,
since it describes the phenomenologically interesting case.
Besides, it makes the conclusions obtained here applicable to the
case studied in \cite{NOIS}, in continuity to the program of
developing formal concepts to braneworld scenarios with torsion. In this
way $D=5$, $p=3$, and consequently $\tilde{R}=0$, because there is
just one dimension on the internal space. With such specifications
and assuming implicitly, as usual, that the brane action volume element
 does not depend on torsion\footnote{In this way, we
guarantee that the brane volume element reduces to $d^{4}x$ in the
limit of null torsion and flat space.}, Eq.(\ref{11}) becomes \ba
\oint
W^{\xi+1}\Bigg(T^{\mu}_{\;\;\mu}+2(\xi-1)T^{m}_{\;\;m}+\frac{\xi}{\kappa_{5}^{2}}\bar{R}W^{-2}\Bigg)=0,\label{12}
\ea where $8\pi G_{5}=\kappa_{5}^{2}=\frac{8\pi}{M_{5}^{3}}$, with
$M_{5}$ denoting the 5-dimensional Planck mass. In order to
implement torsion terms in our analysis, the expressions for the
Riemann and Ricci tensors  in terms of contortion components
related with their partners --- constructed with the usual metric
compatible Levi-Civita connection \cite{AORA}
\begin{eqnarray}
\bar{R}^{\lambda}_{\;\;\tau\alpha\beta}&=&\mathring{\bar{R}^{\lambda}}_{\;\tau\alpha\beta}+\nabla_{[\alpha}K^{\lambda}_{\;\;\tau\beta]}
+ K^{\lambda}_{\;\;\gamma[\alpha}K^{\gamma}_{\;\;\tau\beta]},
\qquad
\bar{R}_{\tau\beta}=\mathring{\bar{R}}_{\tau\beta}+\nabla_{[\lambda}K^{\lambda}_{\;\;\tau\beta]}+K^{\lambda}_{\;\;\gamma\lambda}K^{\gamma}_{\;\;\tau\beta}
-K^{\lambda}_{\;\;\tau\gamma}K^{\gamma}_{\;\;\lambda\beta}
\label{c5}
\end{eqnarray}
are used, where $\nabla$ precisely denotes the covariant derivative without
torsion. The brane scalar curvature is written in terms of the
usual scalar curvature, $\mathring{\bar{R}}$ (without torsion),
plus 4-dimensional contortion terms corrections \cite{AORA} \ba
\bar{R}=\mathring{\bar{R}}+2D^{\lambda}\,^{(4)}\!K^{\tau}_{\;\lambda\tau}-
^{(4)}\!\!K_{\tau\lambda}^{\;\;\;\lambda}\,^{(4)}\!K^{\tau\lambda}_{\;\;\;\lambda}+
^{(4)}\!\!K_{\tau\gamma\lambda}\,^{(4)}\!K^{\tau\lambda\gamma},\label{13}
\ea where the label ``${\scriptstyle{(4)}}$'' on the contortion terms denotes the
contortion of the 3-branes and the covariant derivative is
considered when  a connection that
presents {\it no} torsion is taken into account. In order to verify the influence of
torsion terms in the scalar curvature within the braneworld
context we shall substitute  Eq.(\ref{13}) into (\ref{12}).
First, however, note that in order to reproduce the observable
universe one can put $\mathring{\bar{R}}=0$ with $10^{-120}M_{Pl}$
of confidence level, where $M_{Pl}$ is the Planck mass. Note
that the observations concerning the scalar curvature are related
to the torsionless $\mathring{\bar{R}}$, not to $\bar{R}$. So,
taking it into account it follows that \be \oint
W^{\xi+1}\Bigg[T^{\mu}_{\;\,\mu}+2(\xi-1)T^{m}_{\;\;m}+\frac{\xi
W^{-2}}{\kappa_{5}^{2}}\Big(2D^{\lambda}\,^{(4)}\!K^{\tau}_{\;\lambda\tau}-
^{(4)}\!\!K_{\tau\lambda}^{\;\;\;\lambda}\,^{(4)}\!K^{\tau\lambda}_{\;\;\;\lambda}+
^{(4)}\!\!K_{\tau\gamma\lambda}\,^{(4)}\!K^{\tau\lambda\gamma}
\Big)\Bigg]=0.\label{14}\ee

In order to proceed with the consistency conditions we specify the
standard {\it ansatz} for the stress-tensor. Assuming that there
are no other types of matter in the bulk, except the branes and
the cosmological constant, we have \cite{LE} \ba
T_{MN}=-\frac{\Lambda}{\kappa_{5}^{2}}G_{MN}-\sum_{i}T_{3}^{(i)}P[G_{MN}]_{3}^{(i)}\delta(y-y_{i}),\label{15}
\ea where $\Lambda$ is the bulk cosmological constant,
$T_{3}^{(i)}$ is the tension associated to the i$^{th}$-brane and
$P[G_{MN}]_{3}^{(i)}$ is the pull-back of the metric to the
3-brane. The partial traces of (\ref{15}) are given by \ba
T_{\;\,\mu}^{\mu}=\frac{-4\Lambda}{\kappa_{5}^{2}}-4\sum_{i}T_{3}^{(i)}\delta(y-y_{i}),
\qquad \text{and}\qquad
T_{\;\,m}^{m}=-\frac{\Lambda}{\kappa_{5}^{2}},\label{16}\ea in
such way that  Eq.(\ref{14}) becomes \ba \oint
W^{\xi+1}\Bigg[\frac{2\Lambda}{\kappa_{5}^{2}}(\xi+1)+4\sum_{i}T_{3}^{(i)}\delta(y-y_{i})-\frac{\xi
W^{-2}}{\kappa_{5}^{2}}\Big(2D^{\lambda}\,^{(4)}\!K^{\tau}_{\;\lambda\tau}-
^{(4)}\!\!K_{\tau\lambda}^{\;\;\;\lambda}\,^{(4)}\!K^{\tau\lambda}_{\;\;\;\lambda}+
^{(4)}\!\!K_{\tau\gamma\lambda}\,^{(4)}\!K^{\tau\lambda\gamma}
\Big)\Bigg]=0.\label{18}\ea As one can see, this formalism can be
applied for a several branes scenario. The number of branes,
nevertheless, is not so important to our analysis. To fix ideas
let us particularize the formalism to the two branes case. Denoting
$T_{3}^{(1)}=\lambda$, the visible brane,
$T_{3}^{2}=\tilde{\lambda}$, and assuming that neither the
cosmological constant nor the branes contortion terms do depend on
the extra dimension, Eq.(\ref{18}) gives \ba 4\lambda
W_{\lambda}^{\xi+1}+4\tilde{\lambda}W_{\tilde{\lambda}}^{\xi+1}+\frac{2\Lambda}{\kappa_{5}^{2}}(\xi+1)\oint
W^{\xi+1}
-\frac{\xi}{\kappa_{5}^{2}}\Big(2D^{\lambda}\,^{(4)}\!K^{\tau}_{\;\lambda\tau}-
^{(4)}\!\!K_{\tau\lambda}^{\;\;\;\lambda}\,^{(4)}\!K^{\tau\lambda}_{\;\;\;\lambda}+
^{(4)}\!\!K_{\tau\gamma\lambda}\,^{(4)}\!K^{\tau\lambda\gamma}
\Big)\oint W^{\xi-1}=0,\label{19}\ea where
$W_{\lambda}=W(y=y_{1})$ and $W_{\tilde{\lambda}}=W(y=y_{2})$. Now
some physical outputs of the general Eq.(\ref{19}) are analyzed,
in order to investigate the viability of braneworld scenarios with torsion.
The first case we shall look at relates a factorizable geometry.
Nevertheless, before going forward, we shall emphasize that if one
implements the torsion null case in  Eq.(\ref{19}), it is easy
to see that for $\xi=-1$ one recovers the well known fine tuning
of the Randall-Sundrum model, i.e., \ba
\lambda+\tilde{\lambda}=0\label{20},\ea as expected.

\subsection{Non-warped compactifications with torsion}

The non-warped case is implemented by imposing $W=1$, working
then in a factorizable spacetime geometry. The general approach on
consistency conditions, as exposed before, allows this
possibility. In this Subsection we are therefore concerned with
the viability of braneworld scenarios in the general scope analyzed in
reference \cite{ARK} and in the presence of torsion. The case we
are going to describe here is not the most interesting. We shall,
however, study a little further this simplified case, since it
can provide some physical insight to the warped case.

From Eq.(\ref{19}), the non-warped case reads \ba
\frac{2\Lambda}{\kappa_{5}^{2}}(\xi+1)V+4\lambda+4\tilde{\lambda}-\frac{\xi
V}{\kappa_{5}^{2}}\Big(2D^{\lambda}\,^{(4)}\!K^{\tau}_{\;\lambda\tau}-
^{(4)}\!\!K_{\tau\lambda}^{\;\;\;\lambda}\,^{(4)}\!K^{\tau\lambda}_{\;\;\;\lambda}+
^{(4)}\!\!K_{\tau\gamma\lambda}\,^{(4)}\!K^{\tau\lambda\gamma}
\Big)=0,\label{2111} \ea where $V$ denotes the ``volume'' of the
internal space. Note that for $\xi=0$, the torsion terms do not
influence the general sum rules in the present case. In fact, for
$\xi=0$ it follows that \ba
\frac{V\Lambda}{2\kappa_{5}^{2}}+\lambda+\tilde{\lambda}=0,\label{22}
\ea which states that, for non-warped branes, it is possible to
exist an AdS$_{5}$ bulk, even for strictly positive tension values
associated with the branes. Another interesting case is obtained
for $\xi=-1$. In such case the bulk cosmological constant is
factored out and consequently \ba
\Big(\,{}^{(4)}\!\!K_{\tau\lambda}^{\;\;\;\lambda}\,^{(4)}\!K^{\tau\lambda}_{\;\;\;\lambda}-
^{(4)}\!\!K_{\tau\gamma\lambda}\,^{(4)}\!K^{\tau\lambda\gamma}-2D^{\lambda}\,^{(4)}\!K^{\tau}_{\;\lambda\tau}
\Big)=\frac{4\kappa_{5}^{2}}{V}(\lambda+\tilde{\lambda}).\label{23}
\ea Note that the left hand side (LHS) of (\ref{23}) can be
interpreted, from Eq.(\ref{13}), as the difference between
$\mathring{\bar{R}}$ and $\bar{R}$. In other words, the LHS of
Eq.(\ref{23}) measures the contribution of the torsion terms to
the brane curvature, i. e., it indicates how much the brane
curvature differs itself from zero, due to torsion terms. So, we
can write schematically \ba
\mathring{\bar{R}}-\bar{R}=\frac{4\kappa_{5}^{2}}{V}(\lambda+\tilde{\lambda}).\label{24}
\ea We see that the effect of the torsion in the brane curvature
is proportional to the branes tension values in the two branes
scenario, but it decreases with the distance between the branes.
Moreover, since $\kappa_{5}^{2}=8\pi G_{5}\sim 1/M_{5}^{3}$, such
an effect is about $1/(VM_{5}^{3}$). Therefore, it indicates the
low magnitude of torsion effects in the braneworld scenario with
large extra transverse dimension, since it is suppressed by the
5-dimensional Planck scale and also by the volume of the internal
space. Obviously, in a braneworld scenario which solves the
hierarchy problem the typical scale of the higher dimensional
Planck mass is of order $M_{5}\sim M_{weak}$ and then, the
suppression due to the internal space volume is attenuated.

\subsection{The warped case}

This is the most general case we consider in this Section. In the
absence of a factorizable geometry, some configurations of the
warp factor may be responsible for the right mass partition in the
Higgs mechanism without the necessity of any additional hierarchy
\cite{RSI}. Starting from the general Eq.(\ref{19}), we shall look
at the most important cases, namely $\xi=-1,0,1$.

For $\xi=-1$ we have \ba
\Big(\,^{(4)}\!\!K_{\tau\lambda}^{\;\;\;\lambda}\,^{(4)}\!K^{\tau\lambda}_{\;\;\;\lambda}-
^{(4)}\!\!K_{\tau\gamma\lambda}\,^{(4)}\!K^{\tau\lambda\gamma}-2D^{\lambda}\,^{(4)}\!K^{\tau}_{\;\lambda\tau}
\Big)=\frac{4\kappa_{5}^{2}}{\oint
W^{-2}}(\lambda+\tilde{\lambda}).\label{25} \ea This is the warped
analogue of Eq.(\ref{23}) with the volume of the internal space
replaced by the circular integral of $W^{-2}$ in the denominator
of the right hand side. The same conclusions as the $\xi=-1$ case
of the previous Subsection still hold, but here we call the
attention to the minuteness of the torsion terms: even
contributing with such low magnitude effect to the brane
curvature, it allows the branes to have both the same sign
associated to their respective tension values.

The bulk spacetime type can be better visualized in the $\xi=0$
case. Since all torsion terms of Eq.(\ref{19}) are factored out,
it follows that \ba \frac{\Lambda}{2\kappa_{5}^{2}}\oint W+\lambda
W_{\lambda}+\tilde{\lambda}W_{\tilde{\lambda}}=0.\label{26}\ea
Therefore, as $\oint W<0$, it is easy to see that if $\lambda
,\tilde{\lambda}>0$ then necessarily $\Lambda>0$ corresponding to
an dS$_{5}$ bulk geometry. Otherwise, being $\lambda
,\tilde{\lambda}<0$ one arrives at an AdS$_{5}$ bulk geometry.

For the $\xi=1$ case, a slight modification of Eq.(\ref{25})
deserves a notification. The implementation of $\xi=1$ in the
Eq.(\ref{19}) results in \ba \frac{\Lambda}{\kappa_{5}^{2}}\oint
W^{2}+\lambda
W_{\lambda}^{2}+\tilde{\lambda}W_{\tilde{\lambda}}^{2}-\frac{V}{4\kappa_{5}^{2}}
\Big(2D^{\lambda}\,^{(4)}\!K^{\tau}_{\;\lambda\tau}-
^{(4)}\!\!K_{\tau\lambda}^{\;\;\;\lambda}\,^{(4)}\!K^{\tau\lambda}_{\;\;\;\lambda}+
^{(4)}\!\!K_{\tau\gamma\lambda}\,^{(4)}\!K^{\tau\lambda\gamma}\Big)
=0.\label{27} \ea Now, isolating the torsion contribution to the
curvature we have \ba
\Big(2D^{\lambda}\,^{(4)}\!K^{\tau}_{\;\lambda\tau}-
^{(4)}\!\!K_{\tau\lambda}^{\;\;\;\lambda}\,^{(4)}\!K^{\tau\lambda}_{\;\;\;\lambda}+
^{(4)}\!\!K_{\tau\gamma\lambda}\,^{(4)}\!K^{\tau\lambda\gamma}\Big)=\frac{4\Lambda}{V}\oint
W^{2}+\frac{4\kappa_{5}^{2}}{V}(\lambda
W_{\lambda}^{2}+\tilde{\lambda}W_{\tilde{\lambda}}^{2}).\label{28}\ea
From  Eq.(\ref{28}) we see that the torsion contribution to the
brane curvature is constrained by the internal space volume,
however terms coming from the warped compactification --- as
$\oint W^{2}$ and $\oint W_{\lambda,\tilde{\lambda}}^{2}$ --- can turn this
contribution more appreciable. In particular, the first term of
the right hand side of (\ref{28}) is the dominant one, since it is
not suppressed by the 5-dimensional Planck scale and it is
multiplied by the bulk cosmological constant. We shall make more
comments about these results in the next Section.

\section{Concluding Remarks and Outlook}

This paper concerns some effects evinced by torsion terms
corrections --- both in the bulk and on the brane. To study a
typical gravitational signature arising from a gravitational
system we performed in Section II the analysis based upon the well
known Taylor expansion tools --- strongly reminiscent of the
assumption of a direction orthogonal to the brane --- of the bulk
metric in terms of the brane metric, taking into account bulk
torsion terms. Our main result is summarized by Eq.(\ref{modif}).
It shows how the bulk torsion terms intervene in the black hole
area, in an attempt to find some observable effects arising from
the torsion properties. Again, its highly non-trivial form can be
better studied in the context of a specific model. It is out of
the scope of this paper, nevertheless we shall point a line of
research in this area. It could be interesting to apply the
results found in this Section to some gravitational systems, in
analogy to what was accomplished in standard braneworld scenarios
(see, for instance, references \cite{merc}).

In order to study the behavior of the brane torsion terms we
extend, in Section III, the braneworld sum rules. It was
demonstrated that the consistency conditions do not preclude the
possibility of torsion on the brane. It was shown, however, that
the torsion effects in the brane curvature are suppressed. It
could, in principle, explain a negative result for experiments
from the geometrical point of view. Just for a comparative
complement, in reference \cite{SEN} the 5-dimensional torsion
field, identified with the rank-2 Kalb-Ramond (KR) field, was
considered in the bulk. It was demonstrated by the authors the
existence of an additional exponential damping for the zero-mode
of the KR field arisen from the compactification of the transverse
dimension. In some sense, our purely geometrical sum rules
complete the analysis concerning the presence of torsion, this
time on the brane.

In this paragraph we would like to call attention for some related
issues appearing in the literature. In \cite{sen0, SEN} it was
shown that in an effective 4-dimensional theory on the visible
brane, the KR field --- as a source of torsion --- is suppressed
when a torsion-dilaton-gravity action in a Randall-Sundrum
braneworld scenario is considered, explaining the apparent
insensitivity of torsion in the brane. It was shown, however, that
even in this case the KR field may led to new signatures in TeV
scale experiments, when a coupling between dilaton and torsion is
taken into account. The warped extra-dimensional formalism points
to the presence of new interactions, of significant
phenomenological importance, between the Kaluza-Klein modes of the
dilaton and the KR field.

Briefly speaking, the results of this paper point to the fact that
the hypothesis of a torsionless brane universe may be based upon a
justified impression, since its effects from the bulk (studied
from a quantum field theory approach) and from the brane (analyzed
via the geometrical sum rules) are suppressed by some damping
factor. We emphasize, however, that in the context above, the
naive estimative of the 4-dimensional torsion effects (\ref{28})
must be complemented by the results of a more specific system,
perhaps coming from the Section II results, via Eq. (\ref{modif}).
Such a characterization may put this gravitational and geometrical
approach in the same level, concerning the brane torsion
phenomenology, as, for instance, the massive spectrum of
5-dimensional KR field signature which can be viewed in a
TeV-scale accelerator \cite{SEN}. In this vein, the torsionless
brane universe may be naturally substituted by a more fidedigne
braneworld scenario that contains torsion, and may be useful to a
more precise description of physical theories.

\section*{Acknowledgments}
J. M. Hoff da Silva thanks to FAPESP-Brazil (PDJ 2009/01246-8) for
financial support.

\end{document}